\documentclass[a4paper,11pt]{article}
\pdfoutput=1
\usepackage{amsmath}
\usepackage{epsfig}
\usepackage{graphicx}

\makeatletter

\@addtoreset{equation}{section}
\makeatother
\newcommand{\eq}{\begin{equation}}
\newcommand{\eqx}{\end{equation}}
\newcommand{\eqs}{\begin{equation*}}
\newcommand{\eqsx}{\end{equation*}}
\newcommand{\eqn}{\begin{eqnarray}}
\newcommand{\eqnx}{\end{eqnarray}}
\newcommand{\bea}{\begin{eqnarray}}
\newcommand{\eea}{\end{eqnarray}}
\newcommand{\alg}{\begin{align}}
\newcommand{\algx}{\end{align}}

\newcommand{\f}[2]{\frac{#1}{#2}}

\def\({\left(}
\def\){\right)}
\def\<{\left<}
\def\>{\right>}
\def\O{{\cal{O}}}

\def\[{\left[}
\def\]{\right]}

\begin{document}

\begin{titlepage}
\vskip1cm
\begin{flushright}
\end{flushright}
\vskip0.25cm
\centerline{
\bf \Large Holographic Micro Thermofield Geometries of BTZ Black Holes} 
\vskip1cm \centerline{ \textsc{
 Dongsu Bak,$^{\, \tt a,e}$  Chanju Kim,$^{\, \tt b}$ Kyung Kiu Kim,$^{\, \tt c}$  Jeong-Pil Song$^{\, \tt d}$} }
\vspace{1cm} 
\centerline{\sl  a) Physics Department,
University of Seoul, Seoul 02504 \rm KOREA}
 \vskip0.3cm
 \centerline{\sl b) Department of Physics, Ewha Womans University,
  Seoul 03760 \rm KOREA}
 \vskip0.3cm
 \centerline{\sl c) Department of Physics, Sejong University, Seoul 05006 \rm KOREA}
 \vskip0.3cm

 \centerline{\sl d) Department of Chemistry, Brown University, Providence, Rhode Island 02912 \rm USA}
 \vskip0.3cm
 
  \centerline{\sl e)
Center for Theoretical Physics of the Universe}
 \centerline{\sl 
 Institute for Basic Science, Seoul 08826 \rm KOREA}
\vskip0.3cm
 
 \centerline{
\tt{(\,\,dsbak@uos.ac.kr,~cjkim@ewha.ac.kr,}} 
 \centerline{
 \tt{~kimkyungkiu@sejong.ac.kr, jeong\_pil\_song@brown.edu}\,\,)}
  \vspace{1cm}

\centerline{ABSTRACT} \vspace{0.75cm} \noindent
{We find general deformations of  BTZ spacetime and identify the corresponding thermofield initial states of  the dual CFT.
We deform the geometry by introducing bulk fields dual to primary operators and find the back-reacted gravity solutions to the quadratic  order of the deformation parameter.
The dual  thermofield initial states can be deformed by inserting arbitrary linear combination of  operators at the mid-point of  the Euclidean time evolution 
that appears in the construction of the thermofield initial states. 
The deformed geometries are dual to thermofield states without deforming the boundary Hamiltonians in the CFT side. We explicitly demonstrate that the AdS/CFT 
correspondence is not a linear correspondence in the sense that the linear structure of  Hilbert space of the underlying CFT is realized nonlinearly in the gravity side.
 We also find  that their Penrose diagrams are no longer a square but elongated horizontally due to deformation. These geometries describe  a relaxation of 
generic initial perturbation of thermal system while fixing the total energy of the system. The coarse-grained entropy grows and the relaxation time scale
is of order ${\beta}/2\pi$. We clarify that the gravity description involves coarse-graining inevitably missing some information of nonperturbative degrees.  
}
\end{titlepage}

\section{Introduction}\label{sec1}
Currently there are many available examples of AdS/CFT correspondence \cite{Maldacena:1997re,Gubser:1998bc,Witten:1998qj}, from which one may study various aspects of gravity
and field theories in a rather precisely defined setup. Numerous aspects of strongly coupled field theories have been understood by studying the 
 bulk dynamics based on the AdS/CFT correspondence. However understanding certain  aspects of gravity system are still lacking, 
which in particular include    degrees behind horizon and  
gravitational  singularities. 

In this note, we focus on the gravity dynamics 
based on the 3d BTZ black hole \cite{Banados:1992wn}/ thermofield double \cite{Takahasi:1974zn} correspondence which was first 
introduced in \cite{Maldacena:2001kr}.  Here we consider  three dimensional case only, which of course can  be generalized to other dimensions.  
An interesting deformation \cite{Bak:2007jm, Bak:2007qw} of thermofield double system has appeared based on the Janus geometries  
\cite{Bak:2003jk, Bak:2007jm}. The deformation makes the systems living in the left and the right boundaries of the BTZ black hole different from each other  
with an exactly 
marginal operator turned on. 
The corresponding black hole solution becomes
time-dependent, which is called as Janus time-dependent black hole (TDBH). The corresponding thermofield initial state of the boundary CFT involves
an Euclidean time evolution $U= e^{-\frac{\beta}{4} H_R}e^{-\frac{\beta}{4} H_L}$ where $H_{L/R}$ is respectively for the Hamiltonian of the left/right 
system and $\beta$ is the inverse of the late-time equilibrium  temperature. Looking at the system from the viewpoint of one boundary, the Janus TDBH solution 
describes thermalization of an initial perturbation of thermal system.  Namely the above deformation brings the system an initially out-of-equilibrium state, which will 
be exponentially  relaxed away by thermalization leading to the equilibrium state. This late time behaviors are basically controlled by the physics of 
quasi-normal modes. 
Thus the late-time regime is in a quasi-equilibrium but, 
in general, the system is not even in a quasi-equilibrium
 under relaxation,
during which the thermodynamic variables such as temperature and free 
energy are not well defined.   

In this note, we shall consider rather generic perturbations of the BTZ geometry in the framework of the AdS/CFT correspondence, for which
the boundary Hamiltonians remain intact. 
The thermofield initial states of the system, however, 
can still  be deformed rather generically, which is 
followed by a time evolution by undeformed Hamiltonians. This will be achieved by inserting an arbitrary linear combination of 
operators at the mid-point of the Euclidean time evolution as $U= e^{-\frac{\beta}{4} H_0}e^{- \sum_I C_I  O_I }\,e^{-\frac{\beta}{4} H_0}$ with $H_0$ denoting 
the undeformed BTZ Hamiltonian. Based on the operator-state correspondence, a rather generic perturbation of thermal system can be achieved.
Namely such states are still particularly entangled from the viewpoint of a two sided observer. 
These out-of-equilibrium perturbations will be exponentially relaxed away in the far future. Thus the deformations are describing thermalization of generic
perturbation of thermal system. We illustrate these using scalar primary
operators dual to bulk scalar fields. Below we shall find the explicit 
solution of the
scalar field to the leading order which takes a rather simple form. We  solve the back-reacted geometries to the quadratic order of the scalar perturbation 
parameter which we take as $\gamma$. 

These geometries have many interesting  applications. These  may be viewed as 
a realization of  micro thermofield deformations of the BTZ geometry.
We argue that the bulk observer of a particular side cannot extract the full microscopic information available in the reduced density matrix of the same side
by studying the perturbative  gravity dynamics including full  back-reactions.  
The micro-geometries are also expected to play an important role in understanding the behind-horizon degrees, which is beyond the scope 
of the present work.

In section 2, we present the three dimensional  AdS Einstein scalar system and the BTZ background. In section 3, we present the perturbation equations 
including the gravity back-reactions to their leading order. We solve these gravity equations for the simplest perturbation of the $m^2=0$ scalar field. 
We analyze
the deformation of the corresponding  Penrose diagram and horizon area. In section 4, we present the field theory description of the above perturbation. 
In section 5, we generalize the above construction to micro-geometries
corresponding to other deformations of thermofield states.  In section 6, 
we describe the bulk dynamics and their decoding problem. Last section is devoted to our concluding remarks. In appendices, we present more examples of gravity solutions
for various scalar perturbations.
 
 {\it Note added:} Upon preparing the submission, there appeared a paper \cite{TAKA}, whose results partially overlap with 
ours in this paper.

\section{Einstein scalar system}\label{sec2}
We begin with the three dimensional Einstein scalar system
\eq
S={1\over 16\pi G}\int d^3 x \sqrt{-g}\left( R + \frac{2}{\ell^2}- g^{ab}\partial_a \phi \partial_b \phi-m^2 \phi^2\right)
\eqx
One may turn on linear 
combination of the above bulk scalar fields or even other bulk fields with 
non-zero spins. Here we shall limit our consideration to the case of scalar fields.
There are also in general interactions between these bulk fields, which we shall  ignore in this note.
The dimension $\Delta$ of the corresponding dual operator is related to the mass by
\bea
\Delta(\Delta-d) =\ell^2 m^2
\eea 
where $d$ is the spacetime dimension of the boundary CFT which equals $2$ for the present case. 
For the $m^2=0$ case, this theory can be fully consistently  
embedded into
type IIB gravity \cite{Bak:2007jm}.
For non-zero $m^2$ that corresponds to integral dimensions, the solution can be consistently embedded into IIB supergravity
only for the leading order  fluctuations including the gravity back reaction. 
Here we 
set the AdS radius $\ell$ to be unity  for simplicity and recover it whenever it is necessary.
The Einstein equation  reads
\eq\label{eqofma}
R_{ab}+\left(\frac{2}{\ell^2}-m^2 \phi^2\right) g_{ab} = \partial_a \phi \partial_b \phi 
\eqx
and the scalar equation of motion  is given by
\eq\label{eqofmb}
\nabla^2 \phi-m^2 \phi=0 
\eqx
Any resulting solutions involving nontrivial scalar field will 
be  deformations of the well known AdS$_3 \times S^3\times M_4$  background where $M_4$ may be either $T^4$ or $K3$
\cite{deBoer:1998kjm}. Thus our construction is based on this full microscopic  AdS/CFT correspondence.


The  BTZ black hole  in three dimensions  
 can be written as
\eq
ds^2= - \frac{r^2-R^2}{\ell^2} dt^2+ \frac{\ell^2}{r^2-R^2} dr^2 +r^2 d\varphi^2 
\label{btz}
\eqx
where the coordinate $\varphi$  is circle compactified with  $\varphi \sim \varphi + 2\pi$. Of course here we 
turn off the scalar field.
Note that the horizon is located at $r=R$.   
The regularity near $r=R$ is ensured if the Euclidean time coordinate $t_E$ has a period $\beta=2\pi \frac{\ell^2}{R}$.
The corresponding Gibbons-Hawking temperature is then 
\eq
T=\frac{R}{2\pi \ell^2} 
\eqx
The mass of the black hole can be identified as
\eq
M= \frac{R^2}{8 G \ell^2}
\eqx
The boundary system is defined on a cylinder
\eq
ds^2_B = -dt^2 + \ell^2 d\varphi^2
\eqx
whose spatial size is given by $L= 2\pi \ell$. The central charge of the boundary conformal field theory is 
related to the Newton constant by
\eq
c=\frac{3\ell}{2 G}
\eqx
Thus the entropy of the system beomes
\bea
S= \frac{2\pi R}{4 G} = \frac{c \pi}{3} T \, 2\pi \ell
\eea
while the energy of the system can be expressed as
\bea
M= \frac{c \pi}{6} T^2 \, 2\pi \ell
\eea
in terms of the quantities of CFT.

\section{Linearized perturbation}\label{sec30}

Introducing  new coordinates $(\tau, \mu, x)$   defined by 
 \bea
\frac{r}{R} &=&  \frac{\cos \tau}{\cos \mu} \cr
 \tanh \frac{t R}{\ell^2} &=& \frac{\sin \tau}{\sin \mu}\cr
 x &=& \frac{R}{\ell} \varphi
 \label{btzcoor}
 \eea
 the BTZ black hole metric (\ref{btz}) can be rewritten as
\bea
ds^2=\f{\ell^2}{\cos^2 \mu} \left[-d\tau^2+ d\mu^2+ \cos^2 \tau dx ^2 \right] 
\eea
Motivated by the form of the above metric, we shall make the following ansatz
\bea
\frac{ds^2}{\ell^2}=  {-d\tau^2 + d\mu^2\over A(\tau,\mu,x)}+{dx^2\over B(\tau,\mu,x)},
\quad \quad \quad \phi=\phi(\tau,\mu,x) 
\eea
which describes general static geometries. 
It is then straightforward to show that the equations of motion (\ref{eqofma}) and (\ref{eqofmb})  reduce to
\eqn
&&({\vec\partial} A)^2+\frac{B}{2A} ({\partial_x} A)^2   -A\, {\vec\partial}^2 A=2A -\ell^2 m^2 A \phi^2- A^2 \,({\vec\partial} \phi)^2+ AB \,({\partial_x} \phi)^2 
\cr
&& 3({\vec\partial} B)^2 -2 B\, {\vec\partial}^2 B+ \frac{6B^3}{A^3}({\partial_x} A)^2 
- \frac{2B^2}{A^2}\( {\partial_x} A {\partial_x} B + 2 B{\partial^2_x} A \)
=\frac{B^2}{A}(8-4\ell^2 m^2 \phi^2-4B (\partial_x \phi)^2) \cr
&& {\vec\partial} B \cdot {\vec\partial} \phi  +2  \frac{B^2}{A^2}{\partial_x} A {\partial_x} \phi -  \frac{B}{A}{\partial_x} B {\partial_x} \phi  -2 B \, {\vec\partial}^2 \phi-  2\frac{B^2}{A} {\partial^2_x} \phi +2\ell^2 m^2 \frac{B}{A} \phi=0 ,
\eqnx
where we introduced the notation
${\vec\partial}=(\partial_{\tau},\partial_{\mu})$ with inner product with metric $\eta_{ij}={\rm diag}(-1,+1)$.
This solves the full equations of motion up to some extra integration constants. Using the remaining components of equations of motion,
these integration constant should be fixed further.  

As a power series in $\gamma$, the scalar field may be expanded as
\eq
\phi(\tau,\mu,\varphi)=\sum^\infty_{n=0}\gamma^{2n+1} \phi_{(2n+1)}(\tau,\mu,\varphi) 
\eqx
where we resume the general dependence on the coordinate $\varphi$.
Then the scalar equation in the leading order becomes
\eq
\tan \mu \,{\partial_\mu} h +\tan \tau \,{\partial_\tau} h   +\, {\vec\partial}^2 h - \frac{\ell^2 m^2}{\cos^2 \mu} \,h
+ \frac{\ell^2}{R^2 \cos^2 \tau}\partial^2_{\varphi}\, h=0
\label{eq000}
\eqx
where $h(\tau,\mu,\varphi)$ denotes $\phi_{(1)}(\tau,\mu,\varphi)$. By separation of variables, one may try the ansatz
$h(\tau,\mu) \cos j \varphi$ and $h(\tau,\mu) \sin j \varphi$ with $j=0,1,2,\cdots$. Here for simplicity, we shall consider only  
the case $j=0$ in which  the above equation becomes
\eq
\tan \mu \,{\partial_\mu} h +\tan \tau \,{\partial_\tau} h   +\, {\vec\partial}^2 h- \frac{\ell^2 m^2}{\cos^2 \mu} \,h=0
\label{eq0}
\eqx
  In the following, we will construct the most general solutions
of this equation for the mass corresponding to integral dimensions.

The leading perturbation
of the metric part  begins at $O(\gamma^2)$ with even powers of $\gamma$ only. Let us organize the series expansions of 
the metric variables by
\eq \label{eqab}
A=A_0 \Big(1+{\gamma^2\over 4}a(\tau,\mu)+O(\gamma^4)\Big), \quad
B=B_0 \Big(1+{\gamma^2\over 4}b(\tau,\mu)+O(\gamma^4)\Big),
\eqx
where
\eq
A_0 =\cos^2 \mu, \quad  B_0 = \frac{\cos^2 \mu}{\cos^2 \tau} 
\eqx
The leading order equations for the metric part then become
\bea
&& -2 a + \cos^2 \mu \, {\vec\partial}^2 a
= + 4 \cos^2 \mu  ({\vec\partial} h)^2+ 4  \ell^2 m^2 \,  h^2,
\label{eq1}
\\
&& \sin 2\mu  \,{\partial_\mu} b +2\cos^2 \mu  \tan\tau  \,{\partial_\tau} b+\cos^2 \mu\, {\vec\partial}^2 b  =+4 a+8\ell^2 m^2  h^2 
\label{eq2}
\eea
These linear partial  differential equations (with the source term), (\ref{eq0}), (\ref{eq1}) and (\ref{eq2}) are of our main interest below.
As we discussed before, this set solves the full equations of motion up to some extra homogeneous solutions. Using the remaining components 
of equations of motion,
these coefficients of extra homogeneous solutions should be fixed further.  In this section, we shall be working in the case of $m^2=0$
for which one has $\Delta =2$ with the simplest solution of (\ref{eq0}).

\subsection{Linearized solution including back reaction}

\begin{figure}[ht!]
\centering  
\includegraphics[height=6cm]{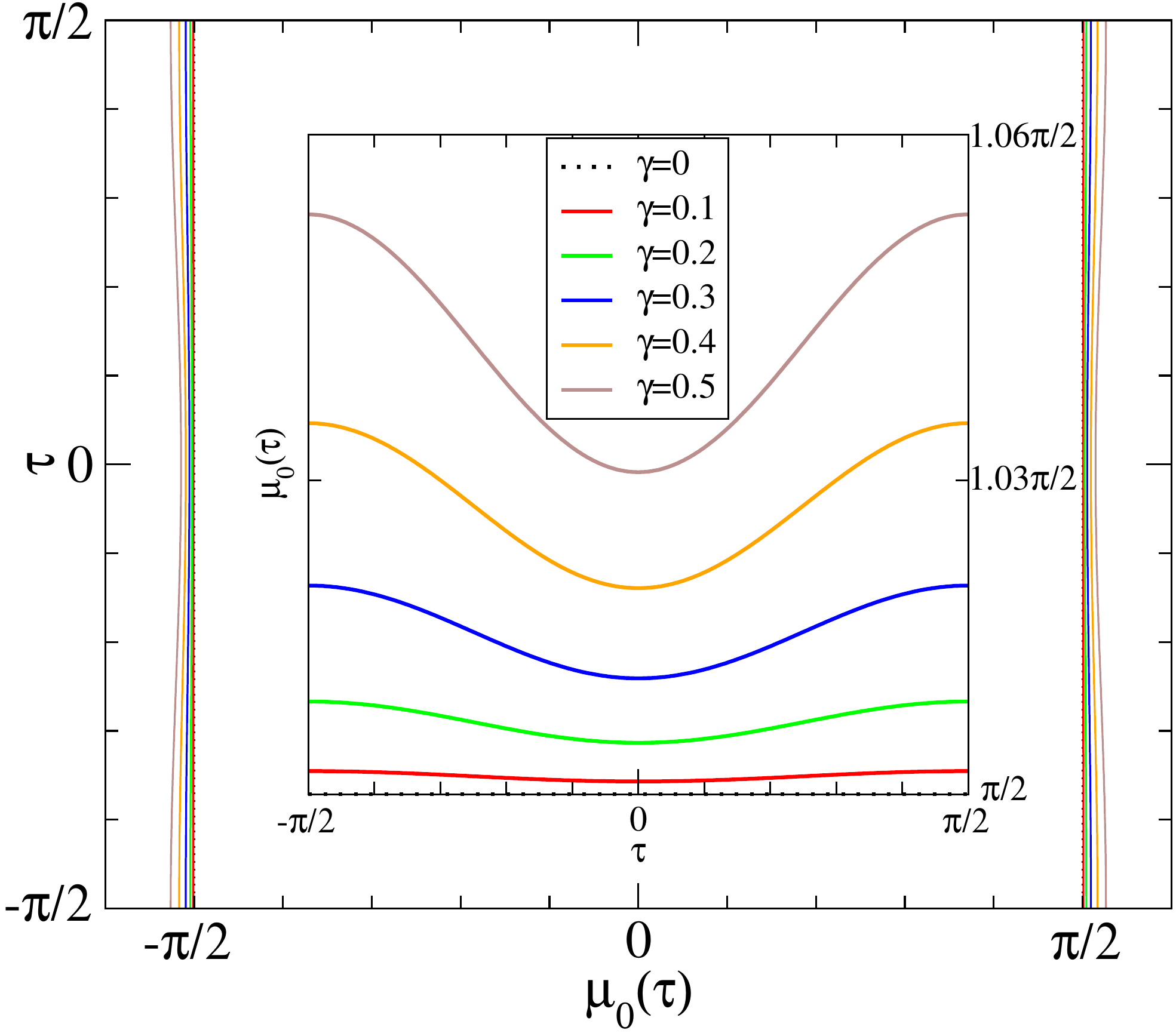}
\caption{Penrose diagram of the perturbed BTZ black hole depicted for various $\gamma$. The relevant  solution involves the scalar perturbation (\ref{simple}) with $m^2=0$.
} 
\label{fig10}
\end{figure}

We begin with a following solution of the leading order scalar equation
\eq
h=\cos^2 \mu \sin \tau 
\label{simple}
\eqx
The solution of \eqref{eq1} and \eqref{eq2}
for the geometry part can be organized as  
 \bea
&& a=\alpha_0 (\mu) + \alpha_1 (\mu) \cos 2\tau\cr
&& b=\beta_0 (\mu) + \beta_1 (\mu) \cos 2\tau
\label{qsolu0}
\eea
where
\bea
\alpha_0 &=& 
 \frac{1}{64} (1 + 6 \cos 2\mu + 5 \cos 4 \mu) + c_1 \tan \mu + 
  \frac{21}{16}(1 + \mu \tan \mu) \cr
\alpha_1 &=& -\frac{1}{16} ( 
    5 + \cos 4\mu + 6 \mu  (2 + \cos 2\mu )  \tan \mu) + 
  c_3\cos^2 \mu  + c_4 (2 + \cos 2\mu )  \tan \mu \cr
 \beta_0 & =& c_2-
 \frac{1}{16} (13 + 16 c_3) \cos^2 \mu  + 
  \frac{3}{8} \cos^4 \mu  + (-2c_4 + 
     \frac{3}{4} \mu ) \cos \mu \sin \mu \cr
&+& 
  \Big(c_1 + \frac{21}{16} \mu \Big)  \tan \mu) \cr
\beta_1 &=& -\frac{1}{32} + \frac{c_3}{2} -\frac{5}{16}  \cos 2\mu   - \frac{3}{32}  \cos 4\mu +
   \Big(c_4 - \frac{3}{8} \mu \Big)  \tan \mu
\eea
We then set all the odd homogeneous terms to zero by requiring 
$c_1=c_4=0$. Then
\bea
\alpha_0 &=& 
 \frac{1}{64} (1 + 6 \cos 2\mu + 5 \cos 4 \mu)+ 
  \frac{21}{16} (1 + \mu \tan \mu) \cr
\alpha_1 &=& -\frac{1}{16} ( 
    5 + \cos 4\mu + 6 \mu  (2 + \cos 2\mu )  \tan \mu) + 
  c_3\cos^2 \mu  \cr
  \beta_0 & =& c_2-
 \frac{1}{16} (13 + 16 c_3) \cos^2 \mu  + 
  \frac{3}{8} \cos^4 \mu  +  
     \frac{3}{4} \mu \cos \mu \sin \mu
+ 
  \frac{21}{16} \mu   \tan \mu \cr
\beta_1 &=& -\frac{1}{32} + \frac{c_3}{2} -\frac{5}{16}  \cos 2\mu   - \frac{3}{32}  \cos 4\mu - \frac{3}{8} \mu   \tan \mu 
\eea
To fix the remaining coefficients $c_2$ and $c_3$, 
now note that the metric functions $A$ and $B$ in \eqref{eqab}
can be written in more convenient forms
\bea
&&\cos^2 \mu\, \(1+ \frac{\gamma^2}{4}(\alpha_0 + \alpha_1 \cos 2\tau)\) = \frac{\cos^2 \kappa \mu}{\kappa^2}  \(1+ \frac{\gamma^2}{4}(\bar\alpha_0 + \bar\alpha_1 \cos 2\tau) + O(\gamma^4)\)\cr
&& \cos^2 \mu \(1+ \frac{\gamma^2}{4}(\beta_0 + \beta_1 \cos 2\tau)\) = \frac{\cos^2 \lambda \mu}{\lambda^2}  \(1+ \frac{\gamma^2}{4}(\bar\beta_0 + \bar\beta_1 \cos 2\tau )+ O(\gamma^4)\)
\label{nonperturbative}
\eea
where we introduce 
\bea
&& \kappa(\tau,\mu) =  1-\frac{\gamma^2}{8} \left( 
\frac{21}{16} -\frac{3}{8}(1+2\cos^2 \mu)\cos 2\tau \right)  + O(\gamma^4)
\cr
&&\lambda(\tau,\mu)= 1-\frac{\gamma^2}{8} \left( 
\frac{21}{16} + \frac{3}{4}\cos^2 \mu  -\frac{3}{8} \cos 2\tau \right)  + O(\gamma^4)
\eea
One then finds 
\bea
\bar\alpha_0 &=&  \frac{1}{64} (1 + 6 \cos 2\mu + 5 \cos 4 \mu)  \cr
\bar\alpha_1 &=& -\frac{1}{16} ( 5 + \cos 4\mu - 6 (2 + \cos 2\mu )  ) +   c_3\cos^2 \mu  \cr
\bar\beta_0 & =& c_2-\frac{21}{16}-\frac{1}{16} (25 + 16 c_3) \cos^2 \mu  +  \frac{3}{8} \cos^4 \mu 
\cr
\bar\beta_1 &=&-\frac{1}{32} + \frac{c_3}{2} -\frac{5}{16}  \cos 2\mu   - \frac{3}{32}  \cos 4\mu  +\frac{3}{8} 
\eea
We now require that $A$ and $B$ have expansions 
\bea
&& \frac{\cos^2 \kappa \mu}{\kappa^2}  \(1+ \frac{\gamma^2}{4}(\bar\alpha_0 + \bar\alpha_1 \cos 2\tau) + O(\gamma^4)\) =(\mu-\mu_0)^2 +O[\, (\mu-\mu_0)^3 ] \cr
&& \frac{\cos^2 \lambda \mu}{\lambda^2}  \(1+ \frac{\gamma^2}{4}(\bar\beta_0 + \bar\beta_1 \cos 2\tau) + O(\gamma^4)\)
=(\mu-\mu_0)^2 +O[\, (\mu-\mu_0)^3 ]
\eea
near infinity with $\mu_0 (\tau)= \frac{\pi}{2\kappa(\tau, \pi/2)}$. 
By comparing the coefficients of $(\mu-\mu_0)^2$, one may fix  
$c_2= \frac{21}{16}$ and $c_3= -\frac98$.
This  choice fixes the freedom of coordinate scaling. 
Therefore one has
\bea
\bar\alpha_0 &=& -\frac{1}{16} \cos^2 \mu ( 7- 10\cos^2 \mu    ) \cr
\bar\alpha_1 &=& \frac{1}{8} \cos^2 \mu ( 1-4\cos^2 \mu   )   \cr
 \bar\beta_0 & =& -\frac{1}{16} \cos^2 \mu ( 7- 6\cos^2 \mu )  \cr
\bar\beta_1 &=&\frac{1}{8} \cos^2 \mu ( 1-6\cos^2 \mu   )  
\eea
Thus
\bea
\alpha_0 &=& 
-\frac{1}{16} \cos^2 \mu ( 7- 10\cos^2 \mu    ) +\frac{21}{16} (1 + \mu \tan \mu) \cr
\alpha_1 &=& \frac{1}{8} \cos^2 \mu ( 1-4\cos^2 \mu   )  -\frac{3}{8}  (1 + 2\cos^2 \mu )(1+\mu   \tan \mu) \cr
 \beta_0 & =& 
-\frac{1}{16} \cos^2 \mu ( 7- 6\cos^2 \mu    )  +\frac{1}{16}  (21 + 12\cos^2 \mu )(1+\mu   \tan \mu) \cr
\beta_1 &=&\frac{1}{8} \cos^2 \mu ( 1-6\cos^2 \mu   ) - \frac{3}{8} (1 + \mu \tan \mu)
\eea
One further finds
\bea
\mu_0 (\tau)=  \frac{\pi} {2}\left[ 1+\frac{\gamma^2}{8} \left( 
\frac{21}{16} -\frac{3}{8} \cos 2\tau \right)  + O(\gamma^4)\right]
\eea
In this coordinate system, the (orbifold) singularity is still located at $\tau=\pm \frac{\pi}{2}$. Hence 
the $\tau$ directional coordinate is ranged over $[-\pi/2, \pi/2]$, which is the same as before. 
On the other hand, the spatial infinity is at $\mu=\pm \mu_0(\tau)$ so that the $\mu$ coordinate is ranged over
\bea
-\mu_0(\tau) \le \mu \le \mu_0(\tau) 
\eea 
We depict the corresponding Penrose diagram of the perturbed BTZ black hole in Fig.~\ref{fig10}. One finds
the Penrose diagram is elongated horizontally in a $\tau$-dependent manner. We find that any boundary two points cannot be
connected by lightlike geodesics through the bulk including the present case as well as the cases discussed below.
This in particular implies that the left and the right boundaries are causally disconnected completely. Hence there 
cannot be any interactions between the left and the right CFT's.  We also find that
$\mu_0(\tau) \ge \pi/2$ for all the cases considered below but we are not so sure if this holds in general.

\subsection{ Boundary stress tensor and horizon area}

We let $O(t, \varphi)$ the operator dual to the scalar field $\phi$. Then its vacuum expectation value may be identified as  
\bea
\langle O(t,\varphi) \rangle = \frac{\gamma R^2}{8\pi G \ell^3}  \frac{1}{\cosh^2 \frac{t R}{\ell^2}}\tanh \frac{t R}{\ell^2}
= \frac{\gamma \,  c \pi }{3\beta^2} 
\frac{1}{\cosh^2 \frac{2
\pi t}{\beta}}\tanh \frac{2\pi t }{\beta}
\label{onep}
\eea 
where we used the standard  holographic dictionary \cite{Skenderis:2002wp}.
This shows exponential decaying behaviors. Here the temperature should be the late time 
equilibrium temperature since the system is time dependent.
The perturbation may be characterized by the initial conditions
\bea
&& \langle O(0,\varphi) \rangle=0 \cr
&&  \frac{\partial}{\partial t}\langle O(t,\varphi) \rangle |_{t=0} =\frac{2c \pi^2}{3\beta^3} 
 \gamma
\eea
The initial perturbation is exponentially relaxed away in late time, which describes a 
thermalization of initial perturbation. The thermalization   is controlled by the time scale 
\bea
t_d =\frac{\beta}{2\pi}
\eea  

Let us now show that the expression for the boundary stress  tensor remains unperturbed. 
For this purpose, we shall construct asymptotic metric which is valid up to order $(\mu-\mu_0)^4$. Let us define
$\bar\mu(\tau)$ by $\bar\mu(\tau)=\mu-\mu_0(\tau)$. The functions $A$ and $B$ can be expanded as
\bea
A &=&    \bar\mu^2 \left(  1-\frac{1}{3}\bar\mu^2  +\frac{\gamma^2}{2}q\, \bar\mu \cos 2 \tau  + \cdots \right)\cr 
B  &=& \frac{\bar\mu^2}{\cos^2 \tau } \left(  1-\frac{1}{3}\bar\mu^2  -\frac{\gamma^2}{2}q\, \bar\mu   + \cdots \right) 
\eea
where 
$q=\frac{3\pi}{16} $ and $\cdots$ denotes higher order terms  in $\bar\mu$ and $\gamma$.
By the coordinate transformation,
\bea
&&\bar\mu =\tilde\mu + \frac{\gamma^2}{4} q \cos 2 \tilde\tau \sin^2 \tilde\mu +\cdots\cr 
&& \tau = \tilde\tau + \frac{\gamma^2}{8} q \sin 2 \tilde\tau \sin 2\tilde\mu +\cdots
\eea
the metric becomes
\bea \label{btzt}
\frac{ds^2}{\ell^2}=  \frac{1}{\tilde\mu^2 \left(  1-\frac{1}{3}\tilde\mu^2 +\cdots \right)}\left[ -d\tilde\tau^2 + d\tilde\mu^2+\cos^2 \tilde\tau \,
dx^2 \right]
\eea
which agrees with the standard BTZ metric. Thus the stress energy tensor remains unchanged.
The mass and pressure are then given by
\bea
M = 2\pi  \ell \, p = \frac{1}{8 G} \frac{R^2}{\ell^2} 
\eea
which are time independent.

\begin{figure}[ht!]
\centering  
\includegraphics[height=4cm]{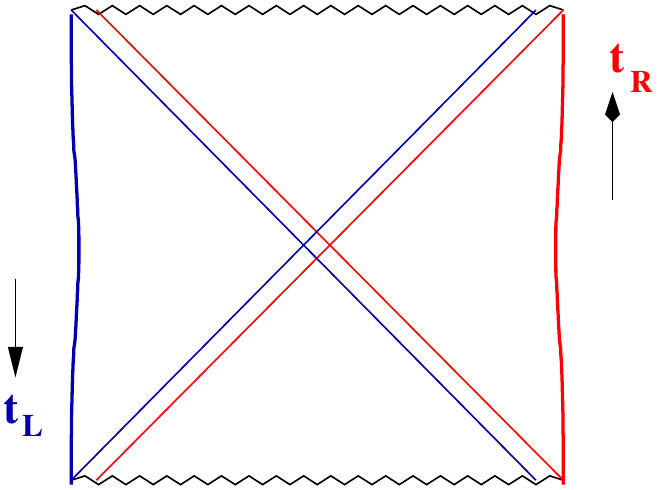}
\caption{The future and past horizons are depicted by straight lines. The horizon length along the future horizon grows monotonically in time. The red lines are the 
horizons of the right side observer
whereas the blue lines represent the horizons of the left side observer.} 
\label{fig2}
\end{figure}

\begin{figure}[t!]
\centering  
\includegraphics[height=5cm]{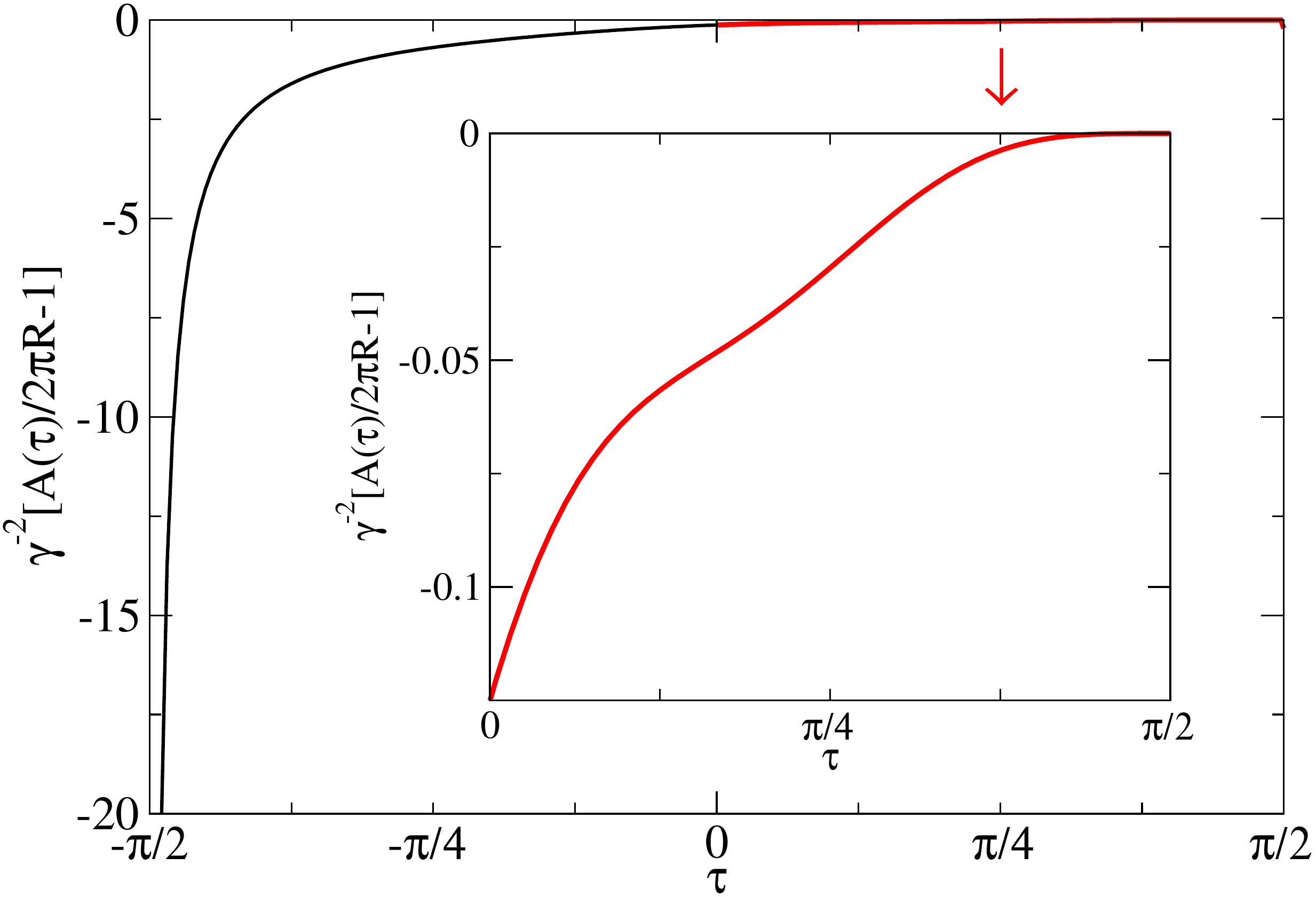}
\caption{The future horizon area minus $2\pi R$ is depicted as a function of $\tau$. 
In the region near $\tau=-\frac{\pi}{2}$, our small $\gamma$ approximation breaks down 
since the coefficient of $\gamma^2$ term becomes too large. The validity requires that $|\mathcal{A}(\tau)/(2\pi R)-1| \ll 1$.} 
\label{fig3}
\end{figure}

In Fig.~\ref{fig2}, we draw the future and past horizons from the both boundaries. The 
horizons associated with the right/left boundary are depicted in red/blue color respectively.
Let us now compute the horizon area along the right-side future horizon that is given by 
\bea
\mu(\tau)= \tau -\frac{\pi}{2}+ \mu_0\Big(\frac{\pi}{2}\Big) 
=\tau +\gamma^2 \frac{27 \pi}{256} + O(\gamma^4)
\eea
The horizon area (length) becomes
\bea
\mathcal{A}(\tau) = 2\pi R \left[ 1- \frac{\gamma^2}{128}\Big(
27 -9 \cos^2 \tau +22 \cos^4 \tau -24\cos^6 \tau  -27\big(\frac{\pi}{2}-\tau\big) \tan \tau
\Big) +O(\gamma^4)\right]
\eea
 In the region near $\tau=-\frac{\pi}{2}$, our small $\gamma$ approximation breaks down 
since the coefficient of $\gamma^2$ term becomes too large. In this region, one has to use 
$B(\tau, \mu)$ in (\ref{nonperturbative}) in the evaluation of $\mathcal{A}(\tau)$, from which one finds
$\mathcal{A}(-\pi/2)=0$ as expected.  We draw the time dependence of the horizon area in Fig.~\ref{fig3}.
One finds $\mathcal{A}(\pi/2)$ agrees with the BTZ value $2\pi R$ whereas
$\mathcal{A}(0)$ is given by
\bea
\mathcal{A}(0)=2\pi R \left[ 1- \frac{\gamma^2}{8} +O(\gamma^4)\right]
\eea
The area is monotonically increasing as a function of time along the future
horizon from zero to $2\pi R$.
The corresponding entropy $S(\tau)=\mathcal{A}(\tau)/4G$ will be interpreted as a coarse-grained entropy of the system as discussed in 
detail in the next section.

\subsection{Convenient form of coordinates}\label{sec33}

One may get a new coordinate system in which  the form of the metric simplifies. For this we make the following coordinate 
transformation
\bea
&& \mu = \sigma + \frac{3\gamma^2}{64} \left( \sigma \cos 2 \sigma
\cos 2\nu  + \frac{7}{2}  \sigma \right)+O(\gamma^4)\cr 
&& \tau = \nu -\frac{3\gamma^2}{64} \left( \sigma  \sin 2 \sigma  + \cos^2
\sigma  +\frac12 \right) \sin 2\nu +O(\gamma^4)
\eea
Then the metric turns into the form 
\bea \label{m0metric}
\frac{ds^2}{\ell^2}=  \frac{1}{\cos^2  \sigma }\left[ 
-\frac{d\nu^2}{ 1+ \frac{\gamma^2}{4} a_\nu+O(\gamma^4) }+\frac{d \sigma ^2}{ 1+ \frac{\gamma^2}{4} a_ \sigma  +O(\gamma^4) }+\frac{\cos^2 \nu \, dx^2}{ 1+ \frac{\gamma^2}{4} b_x +O(\gamma^4) }\right] 
\eea
where
\bea
&&a_ \sigma  = -\frac{1}{16} \cos^2  \sigma  ( 7- 10\cos^2  \sigma    ) - \frac{1}{8} \cos^2  \sigma  ( 11+4\cos^2  \sigma  ) \cos 2 \nu \cr
&& a_\nu  =\frac{21}{16} -\frac{1}{16} \cos^2  \sigma  ( 7- 10\cos^2  \sigma    ) +\frac{1}{8} \cos^2  \sigma  ( 1-4\cos^2  \sigma   ) \cos 2 \nu \cr
&& b_x =\frac{18}{16} -\frac{1}{16} \cos^2  \sigma  ( 1- 6\cos^2  \sigma    ) +\frac{1}{16} \left( -3 +4\cos^2  \sigma  ( 2-3\cos^2  \sigma   ) \right) 
\cos 2 \nu 
\eea
By further coordinate transformation, one may put $b_x$ to zero but there seems no essential simplification in doing so. Note also that the entire Penrose diagram is covered by the coordinate ranges
$\nu, \sigma \in [-\pi/2, \pi/2]  $.

\section{Field theory construction}

\begin{figure}[ht!]
\centering  
\includegraphics[height=4cm]{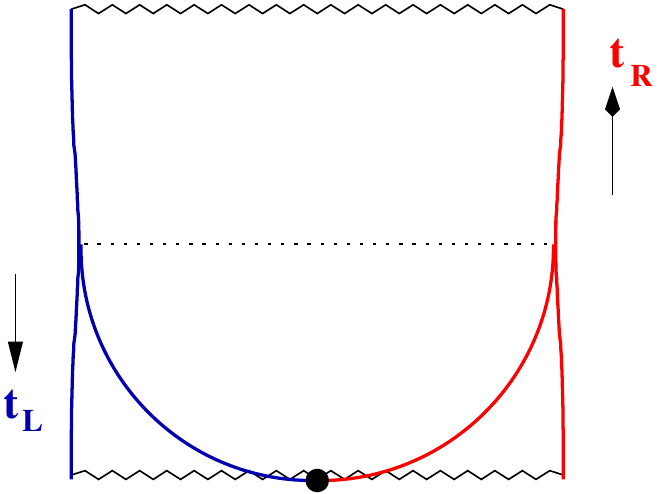}
\caption{The combination of the Lorentzian and Euclidean geometries is depicted. This is used to construct a thermofield initial state and  subsequent 
Lorentzian time evolution. } 
\label{figtdbh}
\end{figure}

In the field theory side,  initial states can be prepared following the thermofield construction in
\cite{Maldacena:2001kr}, which is generalized in \cite{Bak:2007qw}.  Let us begin with general construction first.
We insert operators along the Euclidean boundary, which deforms the field-theory Lagrangian by 
\bea
{\cal L}(-i t_E,\varphi)= {\cal L}_0(-i t_E,\varphi) +\gamma g(t_E, \varphi) O (-i t_E, \varphi) 
\eea
where $t_E = i t$ is the Euclidean boundary time that is circle compactified by
\bea
t_E \sim t_E + \beta
\eea
Here we choose $t_E$ ranged over $[-\beta/2, \beta/2)$ and $g(t_E, \varphi)$ to satisfy the reflection positivity
defined by
\bea
g^*(t_E, \varphi) =g(-t_E, \varphi)
\eea
assuming 
\bea
O^\dagger (t, \varphi) = O (t, \varphi) 
\eea
The Euclidean Lagrangian density is not real in general but the Euclidean action is real.   
Let $H(t_E)$ denote a corresponding Hamiltonian at Euclidean time $t_E$. Then the thermofield initial state is given by
\bea
| \psi (0,0)\rangle = \frac{1}{\sqrt{Z}}\sum_{mn} \langle n | U |  m \rangle\, \, | \bar{m} \rangle_L \otimes | n \rangle_R
\eea
where $Z$ is the normalization factor and $| \bar{m} \rangle$ denotes the state dual to $|m \rangle$.
 The operator $U$ is in general given by
 \bea \label{evolution}
U = T  {\rm exp} \left[ -\int^{0}_{-\frac{\beta}{2}} dt_E  H(t_E) \right]    
\eea 
The Lorentzian  time evolution is given by the Hamiltonian
\bea
-H^T_L (t_L) \otimes 1\, dt_L + 1\otimes H_R (t_R) \, dt_R 
\eea
where the left-right Hamiltonians are identified with
\bea
&& H_L(t_L) =H\Big(\!\! -it_L-\frac{\beta}{2} \Big) \cr
&& H_R(t_R) =H\Big(it_R \Big) 
\eea
We associate the interval $ [-\frac{\beta}{2},-\frac{\beta}{4})$$\oplus$$(\frac{\beta}{4},\frac{\beta}{2}]\,\, /   (-\frac{\beta}{4}, 
\frac{\beta}{4})$ to the 
Lorentzian time $t_L/t_R$ of the left/right system by the 
analytic continuation where $t_L/ t_R$ is ranged over $(-\infty, \infty)$. This is depicted in the Fig.~\ref{figtdbh} where 
both the Euclidean and the Lorentzian geometry appear at the same time. In this figure, we draw only the lower half of Euclidean evolution
which is relevant to the initial ket state. 

The red color is for the right side whereas the blue is for the left system. 
One can motivate the above choice in the following manner. Even with deformations, the coordinate transformation like (\ref{btzcoor}) can be introduced
for the asymptotic regions of the right and the left infinities. Then with $\mu=\pm \mu_0(\tau)$, one has the relations
\bea
&& \tanh \frac{2\pi}{\beta} t_R = \sin \tau \cr
&& \tanh \frac{2\pi}{\beta} t_L =-\sin \tau 
\eea
which identifies the boundary times $t_R$ and $t_L$. Since $\tau$ is ranged over $[-\frac{\pi}{2}, \frac{\pi}{2}]$, one sees that 
$t_R$ and $t_L$ are ranged over   $(-\infty, \infty)$ as expected. Now by analytic continuation, the above becomes
\bea
&& \tan \frac{2\pi}{\beta} t_E^R = \sinh \tau_E \cr
&& \tan \frac{2\pi}{\beta} t^L_E =-\sinh \tau_E 
\eea 
where, from the Euclidean geometry, one finds that $\tau_E$ is ranged over  $(-\infty, \infty)$.  One finds that 
$t_E^R$ can be chosen to be ranged over $(-\frac{\beta}{4}, 
\frac{\beta}{4})$ whereas $t^L_E$ to be ranged over  $[-\frac{\beta}{2},-\frac{\beta}{4})$$\oplus$$(\frac{\beta}{4},\frac{\beta}{2}]$. 
The right and the left parts cover the entire thermal circle in the end. Note
that the points $t_E =\pm\frac{\beta}{4}$ is not associated with 
the right nor the left 
boundaries of the Lorentzian spacetime.  Below we shall use these points to generate the state deformation without deforming the Hamiltonian.

 As we already indicated, the identification of the Lorentzian Hamiltonian involves an analytic continuation from the Euclidean space.  The lower half of the 
 Euclidean solution covered by
 the interval $[-\frac{\beta}{2}, 0]$ is used to construct the thermofield initial state. Then the upper half is associated with the dual state of  
 the thermofield state.
This analytic continuation may not be allowed  in general
unless there is a further restriction on the form of $g(\tau_E, \varphi)$. For the Janus deformation in \cite{Bak:2007qw}, one finds the analytic continuation 
indeed works. We leave further clarification of this issue to future works.  

Thus the time evolution is given by
\bea
| \psi (t_L,t_R)\rangle  = T {\rm exp} \left[ i\int^{t_L}_0 dt'_L  H^T_L(t'_L) \otimes 1   \right]  
 T {\rm exp} \left[ -i\int^{t_R}_0 dt'_R  1 \otimes H_R(t'_R)  \right]  | \psi (0,0)\rangle 
\eea 

With this preliminary, let us consider an entanglement between the left and the right. For this, we introduce so called a reduced density matrix
$\rho_R (t_R)$ defined by
\bea
\rho_R (t_R) = {\rm tr}_L | \psi (t_L,t_R)\rangle   \langle   \psi (t_L,t_R) |
\eea 
where we trace over the left side Hilbert space.
Then the entanglement entropy is defined by the von Neumann definition
\bea \label{srt}
S_R (t_R) = -  {\rm tr}_R \rho_R (t_R) \log \rho_R (t_R)
\eea
This is in general time independent  since $\rho_R (t_R)$ is related to $\rho_R(0)$  by ${\cal U} \,  \rho_R(0) \, {\cal U}^{\dagger}$ with a unitary operator 
${\cal U}= T {\rm exp} \left[ -i\int^{t_R}_0 dt'_R 
H_R(t'_R)  \right] $. 
For the undeformed case with Hamiltonian $H_0$, 
one has $U= e^{-\frac{\beta}{2} H_0}$, which leads to 
$\rho_R = e^{-\beta H_0}/Z_0$. 
One gets  the usual equilibrium thermodynamic entropy out of the entanglement 
 entropy, which is describing the maximal entanglement of the left-right
 systems for a given temperature.  The time-independence of
 \eqref{srt} reflects that the right system alone evolves
unitarily. Thus fine-grained information is fully preserved, which implies that the corresponding fine-grained (von Neumann) entropy  should be time independent.   
In the above, however, we find that the horizon area grows. We interpret the corresponding horizon entropy as a coarse-grained entropy where the coarse-graining may be done 
by ignoring higher-order stringy interactions. In other words, there is a natural coarse-graining due to the gravity approximation that involves the small $G$ 
(or large $c$) limit where especially the nonperturbative degrees are completely missing.  These nonperturbative degrees include those of branes and various nonperturbative 
objects in string theory.
In quantum field theory on $R \times S^1$, one may prove that there is a quantum Poincare  recurrence theorem \cite{Dyson:2002pf} saying that any initial vacuum expectation 
value of any operator should return within a Poincare recurrence time scale.  Our gravity results violate the theorem, which is basically due to the above gravity approximation of 
large $c$ limit.  The fine-grained information is, of course, fully preserved and the coarse-graining due to the approximation is responsible for the violation of the theorem.

An expectation value obtained by insertion of the operator
$O(t,\varphi)$ to the right side boundary is given by
\bea
\langle O(t, \varphi) \rangle=         \langle \psi(0,0)  |  1 \otimes O(t,\varphi)          \,         | \psi(0,0) \rangle
\eea
(Of course one may introduce a one-point function from the left boundary as well.)
This can be evaluated perturbatively as 
\bea
\langle O(t, \varphi) \rangle = \gamma \ell  \int^{\frac{\beta}{2}}_{-\frac{\beta}{2}}ds\int^{2\pi}_{0} d \varphi' g(s, \varphi')
\frac{1}{Z_0} {\rm tr} O(t, \varphi) O(-i(s-\pi), \varphi') e^{-\beta H_0 } + O(\gamma^3)
\label{onep3}
\eea
where $H_0$ is the undeformed CFT  Hamiltonian.
The two-point correlation function is given by
\begin{align} 
\frac{1}{Z_0} {\rm tr} O(t, \varphi) O(t', \varphi') e^{-\beta H_0 } 
\nonumber \\
& \hspace{-41mm}= \frac{\ell^{d- 1} d \, \Gamma(\Delta )\left(\frac{\sqrt{2}\pi}{\beta}\right)^{2\Delta}}{8 \pi^{\frac{d+2}{2}} G \, \Gamma\big(\Delta -\frac{d}{2}\big)}   \sum^\infty_{m=-\infty}\frac{1}{
\left[- \cosh \frac{2\pi}{\beta}( t - t')  + \cosh\frac{2\pi \ell}{\beta} (\varphi-\varphi'+2\pi m)+ i \epsilon\right]^{\Delta}} 
\end{align}
in the AdS/CFT limit. Namely the expression is not exact but involves the large $c$ gravity approximation.
 See \cite{Bak:2007qw, Bak:2015jxd, Bak:2017rpp} for the normalization factor.

For the current problem, we consider the perturbation where $g(t_E, \varphi) 
$ is independent of $\varphi$ with $\Delta =2$.  The one point function then
becomes
\bea 
\langle O(t, \varphi) \rangle  = \frac{\gamma\ell }{8 \pi^2 G} \left(\frac{2\pi}{\beta}\right)^3
\int^{\frac{\beta}{2}}_{-\frac{\beta}{2}}ds\int^\infty_0 dx\,  g_0 (s) \frac{1}{
\left[ -\cosh \frac{2\pi}{\beta}( t+is) +\cosh x\right]^2} +O(\gamma^3)
\label{onep417}
\eea
where $g(t_E, \varphi)=g_0 (t_E)$.
We compare this with the gravity computation in {(\ref{onep})}. 
Thus $g_0(s)$ can be determined by demanding
\bea
\int^{\pi}_{-{\pi}}du \int^\infty_0 dx\,  g_0 \left( \frac{\beta u}{2\pi}\right) \frac{1}{
\left[ -\cosh (v+iu) + \cosh x\right]^2} = \frac{\pi}{\cosh^2 v} \tanh v
\eea
The function $g_0(z)$ is identified  as
\bea \label{gfunction}
g_0(z)= -i \left[ \delta \Big(
\frac{2\pi z}{\beta} -\frac{\pi}{2}
\Big) -  \delta \Big(
\frac{2\pi z}{\beta} +\frac{\pi}{2}
\Big) 
\right] 
\eea
which leads to $H_L (t)= H_R(t)=H_0 $ that is the undeformed
CFT Hamiltonian. 
In other words, the Hamiltonians remain intact under the perturbation 
\eqref{gfunction} which inserts the operator precisely at 
$t_E= -\frac{\beta}4$. (There is, however, an example where the Lorentzian 
Hamiltonians are deformed \cite{Bak:2007qw}.)
Thus the Lorentzian evolution of the thermofield states simplifies as
\bea
| \psi (t_L,t_R)\rangle  = e^{ i H_0 \otimes 1 \, t_L - i 1 \otimes H_0 \, t_R }
| \psi (0,0)\rangle 
\eea 
On the other hand, the thermofield initial state 
$| \psi (0,0)\rangle$ is deformed
because the operator $U$ in \eqref{evolution} is modified to
\bea
U=e^{-\frac{\beta}{4}H_0} \, e^{i\gamma O_{200c}} \,    e^{-\frac{\beta}{4}H_0} =
 e^{-\frac{\beta}{4}H_0} \,\[ 1+  i\gamma O_{200c}+ \O(\gamma^2)  \] \,    e^{-\frac{\beta}{4}H_0}
\eea
where $O_{\Delta njc}$ ($ j\ge 0$) and  $O_{\Delta njs}$  ($ j\ge 1$) are defined by
\bea
&& O_{\Delta njc}= \frac{\ell\beta}{2\pi } \int^{2\pi}_{0} d \varphi \, \cos j \varphi  \left(\frac{\beta}{2\pi }\frac{\partial}{\partial t}\right)^n 
O_{\Delta}(t,\varphi)|_{t=0} 
\cr
&& O_{\Delta njs}= \frac{\ell\beta}{2\pi } \int^{2\pi}_{0} d \varphi \, \sin j \varphi  \left(\frac{\beta}{2\pi }\frac{\partial}{\partial t}\right)^n O_{\Delta}(t,\varphi)|_{t=0} 
\eea
Of course these kinds of definitions may be extended to arbitrary  spin primary operators.

It is clear that an operator insertion at $t_E =-\frac{\beta}{4}$ 
creates a deformation of state without deforming the left and the right 
Hamiltonians.
In fact one may insert an arbitrary linear combination of operators 
\bea
V = \sum_I C_I O_I
\eea
where $O_I$ denote arbitrary linearly independent operators. 
Based on the operator state correspondence,  this leads to rather general deformation of states
without deforming the Hamiltonians of the system.  In the next section we shall illustrate gravity solutions corresponding to 
such deformation of states by the above operators.

\section{Other examples of micro-geometries} 

Other perturbation can be generated in many ways. Here we are interested only in the case where the boundary  Hamiltonians are undeformed as in the previous section.
One way to generate such perturbation is to choose $g(s,\varphi)=g_n(s)$ with
\bea
g_n(s)= \left(i \frac{\beta}{2\pi }\frac{d}{ds}\right)^n  g_0(s)
\eea
The corresponding  expectation value can be given by
\bea
\langle O(t,\varphi) \rangle_n = \left(\frac{\beta}{2\pi }\frac{\partial}{\partial t}\right)^n \langle O(t,\varphi) \rangle_0 + \O(\gamma^3)
\eea
which is derived from the formula (\ref{onep3}).
The scalar field solution can be generated similarly by
\bea
h_n (\tau,\mu)= \left( \frac{\beta}{2\pi }\frac{\partial}{\partial t}\right)^n h_0(\tau,\mu)
\eea
where the subscript $0$ refers to our solution in  section \ref{sec30}. 
This formula partly follows from the fact that
the linearized scalar equation in (\ref{eq000}) involves 
only coefficients which are independent 
of $t$ when the equation is written in terms of coordinates $(t, r,x)$. 
Thus partial derivatives with respect to $t$ generate
new solutions of the linearized equation in \eqref{eq000}. 
For $n=1$ case, one finds from the relation (\ref{btzcoor}) that
\bea
h_1(\tau,\mu)= \gamma \cos^2 \mu \sin \mu (1- 3 \sin^2 \tau)
\eea
and
\bea
\langle O(t,\varphi) \rangle_1 
=\gamma \frac{c}{12\pi} \frac{ R^2}{\ell^4}  \frac{1}{\cosh^2 \frac{t R}{\ell^2}}\left[ 
-2 + \frac{3}{\cosh^2 \frac{t R}{\ell^2}} 
\right]
\label{onep1}
\eea
The analysis of the corresponding back-reacted geometry is presented in Appendix \ref{appa}.
We obtain the deformation of the Penrose diagram which is again elongated horizontally. All the features of this solution
are basically similar to those of the previous solution. In particular, this again describes the physics of 
thermalization though the detailed functional form is different from that of the previous solution. 

\begin{figure}[ht!]
\centering  
\includegraphics[height=5cm]{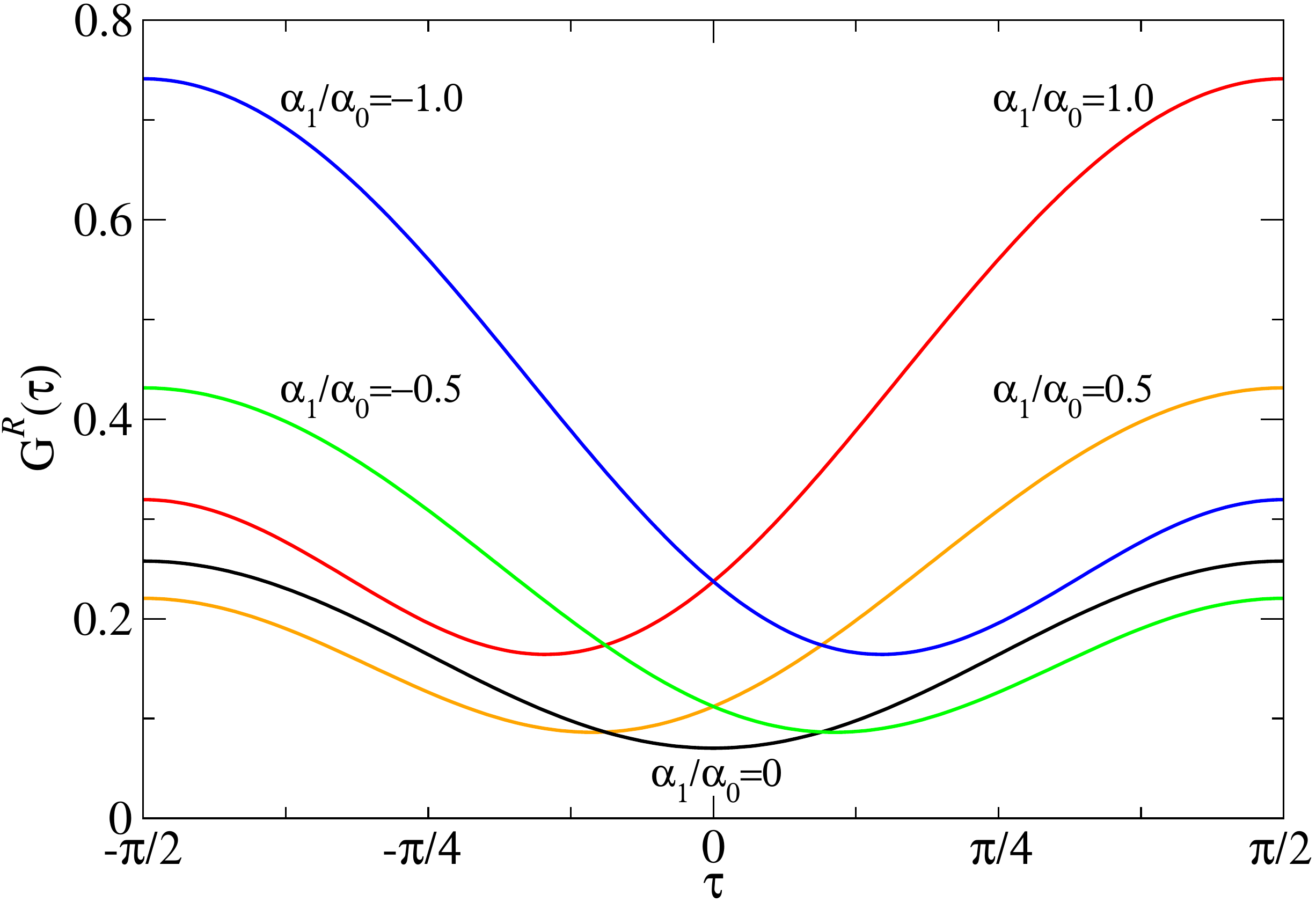}
\caption{We depict here the boundary shapes given by $G^R(\alpha_0,\alpha_1)$ of the Prenrose diagram of the spacetime  with the linear combination 
of $h_0$ and $h_1$. We set $\alpha_0=1$ in the figure. The shape of the left side is given by the relation $G^L(\alpha_0,\alpha_1)=G^R(\alpha_0,-\alpha_1)$.} 
\label{figlin}
\end{figure}

Let us now consider an arbitrary  linear combination of $h_0$ and $h_1$. Namely the linear combination 
\bea
h (\tau,\mu)= \alpha_0\, h_0(\tau,\mu) + \alpha_1 \, h_1(\tau,\mu)
\eea
solves the linearized scalar field equation in (\ref{eq0}) where $\alpha_0$ and $\alpha_1$ are real. From this one may solve the linearized Einstein equations. 
Fortunately we do not have to solve the problem from the beginning. One finds
\bea
&& a (\tau,\mu) =   \alpha^2_0 \, a_{200}(\tau,\mu) + \alpha^2_1 \, a_{210}(\tau,\mu)  +\alpha_0 \alpha_1 \, a_{201}(\tau,\mu) \cr
&& b (\tau,\mu) =   \,\alpha^2_0 \, b_{200}(\tau,\mu)+ \alpha^2_1 \, b_{210}(\tau,\mu)  +\, \alpha_0 \alpha_1 \, b_{201}(\tau,\mu)
\label{nonlinear}
\eea 
where we use the notation $f_{\Delta n_1 n_2}(\tau,\mu)$. Here $n_2=0$ denotes that the solution of linearized Einstein equations is obtained
with the scalar solution $h_{n_1}(\tau,\mu)$. On the other hand,  the nonvanishing $n_2$ implies 
\bea
 f_{\Delta n_1 n_2}(\tau, \mu)= \left( \frac{\beta}{2\pi }\frac{\partial}{\partial t}\right)^{n_2}  f_{\Delta n_1 0}(\tau, \mu)
\eea
Thus the cross terms  follow from $a_{200}$ and $b_{200}$ by simply taking a derivative $\frac{\beta}{2\pi }\frac{\partial}{\partial t}$, 
which one may verify directly by solving the full equations of motion and 
fixing the homogeneous solutions. 
From the solution, again one can work out the field theory implications which are straightforward.  Here let us just mention 
the shape of Penrose diagram which is dictated by $\mu^R_0(\tau)$ and $\mu^L_0(\tau)$  where
$\mu$ is ranged over $[-\mu^L_0(\tau), \mu^R_0(\tau)]$. One finds that
\begin{align}
\mu^{R/L}_0(\tau) = \frac{\pi}{2\kappa(\tau, \pi/2)}
     = 1+ \gamma ^2 G^{R/L}(\tau) + O(\gamma^4)
\end{align}
with
\bea
 G^{R/L}(\tau) &=& \frac{\alpha_0^2}{8} \left( 
\frac{21}{16} -\frac{3}{8} \cos 2\tau \right) + \frac{3\alpha_1^2}{1024}   (74 - 18 \cos 2\tau + \cos4\tau )\cr
&\pm& \frac{3\alpha_0 \alpha_1}{128} \big(10 \sin \tau + \sin 3 \tau\big)
\eea
where $\pm$ are  for $R$ and $L$ respectively.
We draw these functions in Fig.~\ref{figlin} to show the changes in the shape of the Penrose diagram. The shapes of the right 
boundary are illustrated for various $\alpha_0$ and $\alpha_1$ with $\alpha_0=1$. The shape of the left side is given by the relation $G^L(\alpha_0,\alpha_1)=G^R(\alpha_0,-\alpha_1)$.  

There are further linearly independent perturbations with $m^2=0$. 
We choose the function $g(s)$ by
\bea
\bar{g}_0(s)= 
\delta \Big(
\frac{2\pi s}{\beta} -\frac{\pi}{2}
\Big) + \delta \Big(
\frac{2\pi s}{\beta} +\frac{\pi}{2}
\Big) 
\eea
The corresponding scalar field reads
\bea \label{scalarm02}
\bar{h}_0 = \frac{2}{\pi}\cos^2 \mu \left(1-\frac{1}{2}\log \left(\frac{1+\sin\tau}{1-\sin\tau}\right)\sin \tau \right) 
\eea
and the  vev becomes
\bea
\langle O(t,\varphi) \rangle 
=\gamma \frac{c}{6\pi^2} \frac{ R^2}{\ell^4}  \frac{1}{\cosh^2 \frac{t R}{\ell^2}}\left[ 
1 - \frac{t R}{\ell^2} \tanh \frac{t R}{\ell^2}
\right]
\label{onep2}
\eea
One may get the back-reacted solution for the gravity part but we find it is too complicated to present.  The choice 
\bea
g(s,\varphi)=\bar{g}_n(s)= \left( \frac{i\beta}{2\pi }\frac{d}{ds}\right)^n  \bar{g}_0(s)
\eea
will also give the scalar solution given by
\bea
\bar{h}_n (\tau,\mu)= \left( \frac{\beta}{2\pi }\frac{\partial}{\partial t}\right)^n \bar{h}_0 (\tau,\mu)
\eea

 Finally we consider the case of massive scalar whose dual operator $O_\Delta$ has a general dimension $\Delta$. 
The scalar equation \eqref{eq0} then has a simple solution in terms of Legendre
functions,
\begin{equation} \label{legendre}
	h = \cos^\Delta\mu\, (\kappa_1 P_{\Delta-1}(\sin\tau) + \kappa_2 Q_{\Delta-1}(\sin\tau) )
\end{equation}
Note that this reduces to \eqref{simple} or \eqref{scalarm02} for massless case ($\Delta=2$).
Here we consider only the case with $\ell^2 m^2=3$ ($\Delta=3$) and $\kappa_2=0$ for which 
the explicit form of the solution is given by
\begin{align}
	h = \cos^3\mu \left(-\frac23 + \cos^2\tau \right)
\end{align}
 We present  the corresponding back-reacted geometry explicitly in Appendix 
\ref{appb}.

Let us now clarify the general structure of the Hilbert space of the boundary field theory and its realization in the gravity 
solution. For any Hermitian operator $O_I$ constructed from some primary operator dual to the corresponding matter field in the gravity 
side, one may construct a rather general state by the insertion
\bea
U=  e^{-\frac{\beta}{4}H_0} \,\[ 1+ \gamma V + \O(\gamma^2)  \] \,
e^{-\frac{\beta}{4}H_0}
\eea
with $V = \sum_I C_I O_I $
where $C_I$ are arbitrary complex numbers.
For instance, for the operator $O_{200}$, one can choose the linear combination
\bea
g(s, \varphi)= \alpha_0\,  g_0(s) + \bar\alpha_0 \, \bar{g}_0(s)
\eea   
which leads to 
\bea 
V  = (\bar\alpha_0 + i \alpha_0) O_{200} = C_{200} \, O_{200}
\eea
where we take $\bar\alpha_I$ to be real. It is clear that the full Hilbert space of the underlying CFT is linearly realized by 
 the inserted operator $V$. The realization of the leading order solution of matter part is unconventional though it is still linear.
Namely one has 
\bea
h = \alpha_0\, h_0 + \bar\alpha_0 \, \bar{h}_0
\eea 
for the above example which does not realize the complex structure of the Hilbert space properly. Further
the back-reaction of the gravity part is essentially nonlinear as is clear from the explicit solution (\ref{nonlinear}).
Hence we conclude that the AdS/CFT correspondence is not a linear correspondence  in the sense that the linear structure of  Hilbert space of the underlying 
CFT is realized nonlinearly in the gravity side. But we would like to emphasize that the gravity solution reflects all those information of the Hilbert space of 
the perturbative gravity description. As we discussed already, the gravity description misses the nonperturbative degrees such as branes and other nonperturbative 
objects in string theory.

\section{Bulk dynamics}
\begin{figure}[ht!]
\centering  
\includegraphics[height=4cm]{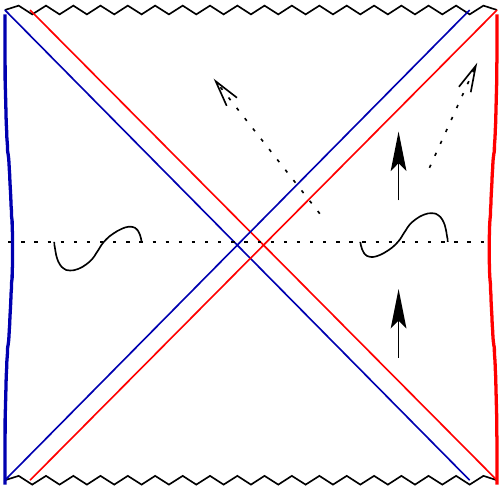}
\caption{We illustrate here  the bulk perturbative dynamics. The initial condition is given at $\nu=0$.  The left and the right initial perturbations 
can be independent from each other.
The dotted lines with an arrow represent possible bulk observer's trajectories. All the information they can gather lies in the right side of horizons that are colored in red.} 
\label{fig90}
\end{figure}

In this section, we shall discuss the behavior of the bulk field based on the above solutions. For an illustration, let us focus on the case of
$m^2=0$ without the angular 
dependence on $\varphi$ with $j=0$. 
The most general solution in the leading order is given by
\bea
h(\tau,\mu) = \sum^\infty_{n=0}\[ \alpha_{n}\, h_{n}(\tau,\mu) + \bar\alpha_n \, \bar{h}_n(\tau,\mu)\]
\eea 
In order to cover the entire Penrose diagram which is deformed by perturbations, it is better to use the coordinates $(\nu, \sigma) \in [-\pi/2, \pi/2]^2$ that 
cover the entire Penrose diagram 
as introduced in section~\ref{sec33}. 
Namely 
\bea
h(\nu, \sigma) = \, h_{e}(\nu, \sigma) +   \, \bar{h}_o(\nu, \sigma) 
\label{wave}
\eea 
with
\bea
h_e(\nu, \sigma) = \sum^\infty_{n=0}  
\bar\alpha_n \, \bar{h}_n(\nu, \sigma)~~,~~
h_o(\nu, \sigma) = \sum^\infty_{n=0}  
\alpha_n \, {h}_n(\nu, \sigma)
\eea 
gives a solution fully covering the deformed Penrose diagram which is also valid to the leading order since the correction due to geometry gives 
 $\O(\gamma^3)$ contributions. 
We shall discuss properties of this solution. First of all, there is a symmetry
\bea
 h_{n}(-\nu, -\sigma) &= &-h_{n}(\nu, \sigma) \cr
 \bar{h}_{n}(-\nu,-\sigma) &= & \bar{h}_{n}(\nu, \sigma) 
\eea
which leads to the symmetry of the solution 
\bea
 h_{o}(-\nu, -\sigma) &= &-h_{o}(\nu, \sigma) \cr
 {h}_{e}(-\nu,-\sigma) &= &{h}_{e}(\nu, \sigma)
\eea
This symmetry basically follows from the symmetry of the BTZ background and our choice of the thermofield initial state. 
The perturbation satisfies the spatial boundary condition $h(\nu,\pm\pi/2)=0$,
which is our choice since there are 
examples \cite{Bak:2007jm,Bak:2007qw,Bak:2011ga} for which 
this condition is relaxed. 
Now we shall give an initial condition at $\nu=0$ by
\bea
 h(0,\sigma) &= & q_1(\sigma )\cr
  \partial_\nu h(\nu, \sigma) |_{\nu=0} &= &q_2 (\sigma) 
\eea
We illustrate this bulk  perturbative dynamics in Fig.~\ref{fig90}, where the left and the right initial perturbations can be independent from each other.
Note that the set  
\bea 
\{h_{2n+1}(0,\sigma)/\cos\sigma, \ \bar{h}_{2n}(0,\sigma)/\cos\sigma  \ | \ n=0,1,2, \cdots \}
\eea
forms a complete basis satisfying Dirichlet boundary condition for the interval $\sigma \in [-\pi/2, \pi/2]$. Hence 
$q_1(\sigma)= \cos \sigma f_1({\sigma})$ where $f_1(\sigma)$ is an arbitrary real function satisfying the Dirichlet boundary condition.
The $\cos \sigma$  factor here follows from the fact that we are considering the bulk field dual to the dimension two
operator. Similarly 
$q_2(\sigma)= \cos \sigma f_2({\sigma})$ where $f_2(\sigma)$ is an arbitrary real function satisfying the Dirichlet boundary condition
where now the basis is given by
 \bea 
\{\partial_\nu h_{2n}(\nu,\sigma) |_{\nu=0}/\cos\sigma, \ \partial_\nu\bar{h}_{2n+1}(\nu,\sigma) |_{\nu=0} /\cos\sigma  \ | \ n=0,1,2, \cdots \}
\eea
 Thus we find that the initial configuration together
with the velocity can be fully localized in the bulk once it satisfies the boundary condition. 
In particular one can choose initial conditions such that it can be fully localized behind the horizon.
The subsequent  $\nu$ development is determined by the wave equation in (\ref{eq0}) which is defined in the fully extended BTZ spacetime. 
The time evolution is well defined except the divergence at the orbifold singularities $\nu= \pm \pi/2$ where $\bar{h}_n$ diverges. 
These are associated with the problems of the singularities behind horizon, to which we have nothing to add in this note. (As will be argued below, our gravity 
description fails near $\tau=\pm \pi/2$ where the singularity is located.) 
Their features are not different from 
those of cosmological singularities in the sense that the singularities are spacelike. Away from $\nu=\pm \pi/2$, its time evolution is ordinary. 
In particular nothing special happens near horizon regions. 

Now we would like to discuss the decoding of information in relation with the above setup. We shall discuss the problem from the 
viewpoint of the observer of the right boundary. 
The information we are interested in is contained in 
the coefficients $(\alpha_n,\bar\alpha_n)$.
There is no subtlety in this bulk description since the simple wave equation governs this reduced  information content of the system.
There are three levels of  available descriptions. First is the description in terms of the solution (\ref{wave})  of the wave equation (\ref{eq0}).   
The second is 
the full gravity description 
allowing back-reactions, which is coarse-grained from the viewpoint of the full microscopic degrees
as we discussed before. 
Finally  there is the full microscopic description by the boundary field theory. In particular $\rho_R(t_R)$ 
contains all those microscopic information available from the viewpoint of
an observer on the right boundary.     
This will require the full string theory from the viewpoint of the bulk. As we demonstrated already, the information contained in $\rho_R$ does not 
change in 
time and hence all the initial information is preserved in time. On the other hand, at the level of gravity description, one finds the future horizon area 
grows which we also demonstrated already.  Hence for the bulk observer staying outside the future horizon, the less region of $\sigma$ is available
observationally. The observer is then able to determine less information on the coefficients $(\alpha_n,\bar\alpha_n)$ as the horizon area grows. 
Hence the information seems to disappear from the bulk observer at this linearized level. The observer may jump into the black hole interior. But he cannot cross 
the past horizon of the right side, which is the $-45^\circ$ red line in Fig.~\ref{fig90}.  Hence one seems to find again less information available 
 since the bigger region is excluded from 
observation. The semiclassical treatment does not help either since the problem is basically from the causality imposed by the horizon. 

Note however that the effect is of order $\gamma^2$ since missing information is mainly due to the horizon change that is of order $\gamma^2$.
The higher order contributions including gravity back-reaction help here which can contain the missing information. 
(If we know all of $(\alpha_n,\bar\alpha_n)$ for instance, 
the higher order contributions give completely redundant information on $(\alpha_n,\bar\alpha_n)$.) There are other ways to argue the recovery 
of information by the higher
order effects. 
One considers the coupling of the left-right boundary by the double trace deformation \cite{Gao:2016bin}. Then this makes Penrose diagram contracted instead of 
elongation which leads to an effective reduction of the horizon area. Hence this way one may recover the missing information at the level of wave equation.    
Therefore all the information regarding the perturbative gravity fluctuation may be restored.  

On the other hand, 
we have  demonstrated  that, within the gravity description,  the expectation value of operator decays exponentially in time violating 
the Poincare recurrence theorem. This in particular implies that the gravity description is not valid at $t = \pm \infty$ (or $\tau=\pm \pi/2$) 
where we set out initial state at $t=0$.  
Thus the missing information at the microscopic level should lie in the degrees that are responsible for the dynamics beyond gravity 
approximation.    
These degrees are coarse-grained within the gravity description.  Their dynamics are nonperturbative in the sense that we do not have a well-defined geometric 
description of micro-geometries.   

This shows that the information loss cannot be resolved within the perturbative gravity framework even if one includes its perturbative back-reactions.  
We do not know how the missing information is stored in such nonperturbative degrees.

\section{Conclusions}

In this note we have considered the deformation of BTZ black holes in the context of AdS/CFT correspondence.  The geometry is dual to a deformation of  
thermofield  initial state while the boundary Hamiltonians remain intact. To deform initial states, we insert a generic linear combination of operators to the mid-point of the 
Euclidean time evolution which is used to construct the thermofield initial states. For each insertion, we can construct the corresponding back-reacted geometries.
The corresponding geometries  encode the information of the CFT side though their relation is highly nonlinear. The resulting geometries describe the exponential relaxation
of any initial perturbation above the thermal vacuum, which is the thermalization of any initial perturbation.  

Our construction of the micro geometries has many potential applications. One may compute for instance multi-point functions from our geometry. Especially evaluation of the 
out-of-time-order 4-point function \cite{Maldacena:2015waa} that shows the quantum chaos behavior \cite{Shenker:2013pqa} is rather straightforward. Here we expect one can compute the behavior of the 
4-point function that is valid for entire range of time without any further restriction. This 4-point function involves an insertion of operators from the both boundaries
 at the same time. One finds that  the behind-horizon degrees are relevant in the evaluation of the 4-point function.
We will report the related study elsewhere.   

Our construction of micro-thermofield geometries are different from the fuzzball proposal \cite{Mathur:2005zp} in many ways. First of all our 
construction is entirely based on the standard AdS/CFT correspondence. Our micro geometries do 
not involve any particular bulk 
local structures on which the fuzzball proposal is based on. Moreover, our deformation always involves black hole horizon
though it is not entirely clear whether the existence of horizon is a necessary condition or not. Of course one still has a pure 
state description from the viewpoint of the total system of the both boundaries. This is sharply contrasted with 
the fuzzball  proposal where the existence of any horizon in the bulk is disputed.

\section*{Acknowledgement}
We would like to thank Hyunsoo Min for his contribution at the early stage of this work. 
D.B. was
supported in part by
NRF Grant 2017R1A2B4003095. 
K.K. was supported by 
NRF  
Grant 
2015R1D1A1A01058220 and by the faculty research fund of Sejong University in 2017.

\appendix

\section{Other perturbation with $m^2=0$}\label{appa}

\begin{figure}[ht!]
\centering  
\includegraphics[height=5.5cm]{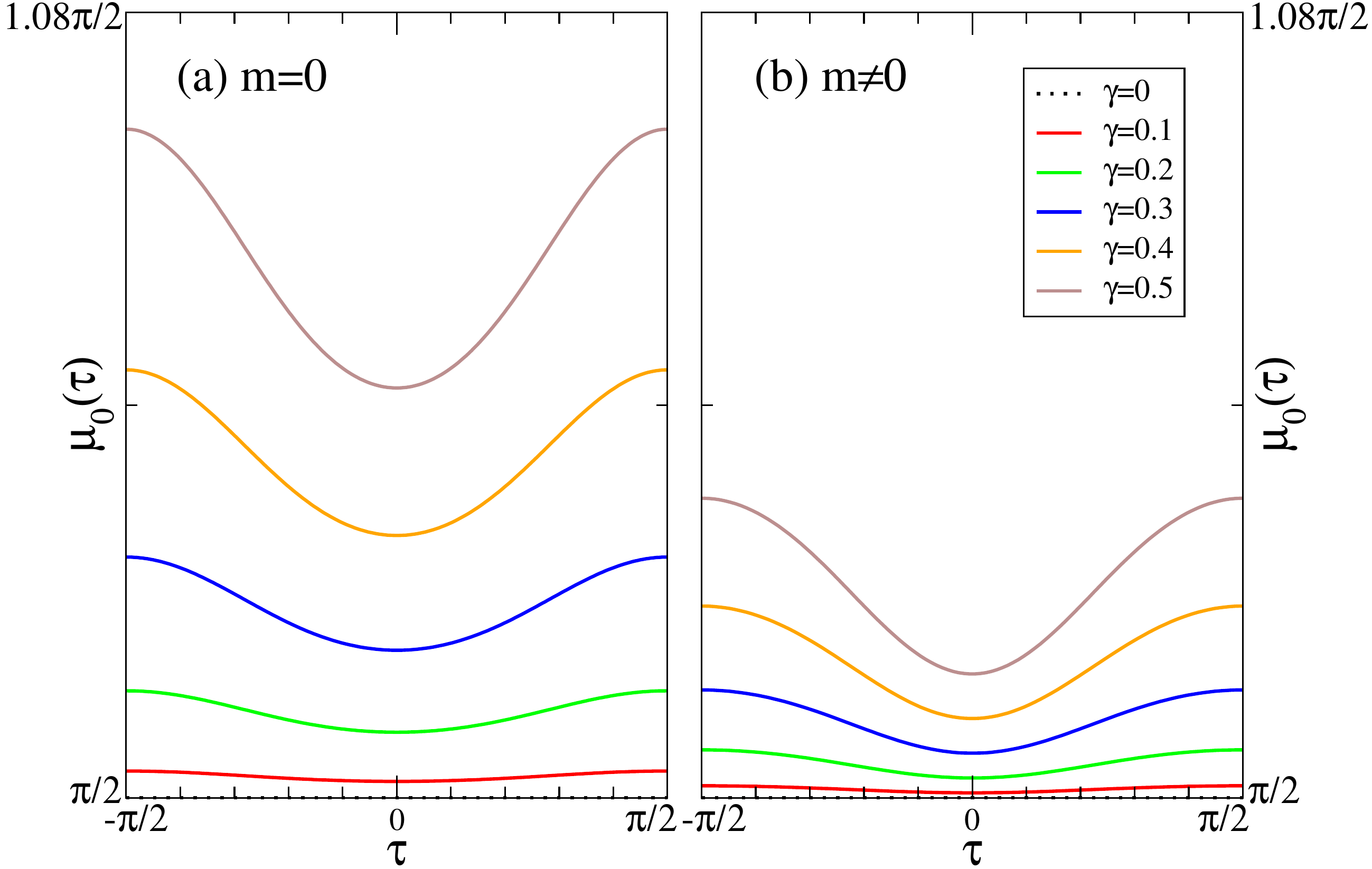}
\caption{Penrose diagrams of the perturbed BTZ black hole. (a) $m=0$ with the
perturbation \eqref{m0h1} (b) $m=\sqrt3/\ell$ with the perturbation
\eqref{m3h}.} 
\label{figcom}
\end{figure}

For $n=1$ case, one has
\begin{equation}\label{m0h1}
h_1 = \gamma \cos^2 \mu \sin \mu (1- 3 \sin^2 \tau)
\end{equation}
Following section 3, we can find the corresponding perturbation for the 
gravity part in the form
\begin{align} \label{m0sol2}
A&=\cos^2\mu \left( 1 + \frac{\gamma^2}4 a \right)
  =\frac{\cos^2\kappa\mu}{\kappa^2} \left( 1 + \frac{\gamma^2}4 \bar a \right)
	= (\mu - \mu_0)^2 + O[(\mu - \mu_0)^3]
\nonumber\\
B&=\frac{\cos^2\mu}{\cos^2\tau} \left( 1 + \frac{\gamma^2}4 b \right)
  =\frac{\cos^2\lambda\mu}{\lambda^2 \cos^2 \tau}
          \left( 1 + \frac{\gamma^2}4 \bar b \right)
	= \frac{(\mu - \mu_0)^2}{\cos^2\tau} + O[(\mu - \mu_0)^3]
\end{align}
where
\begin{align} \label{m0sol2ab}
a&=\alpha_0(\mu) + \alpha_1(\mu) \cos 2\tau
	+ \alpha_2(\mu) \cos 4\tau \\ \nonumber
b&=\beta_0(\mu) + \beta_1(\mu) \cos 2\tau + \beta_2(\mu) \cos 4\tau
\end{align}
with
\begin{align}
\alpha_0 &= \frac1{64} (111 -37 \cos ^2\mu  + 126 \cos^4\mu -120 \cos^6\mu)
	+ \frac{111}{64} \mu  \tan \mu \nonumber\\
\alpha_1 &= -\frac3{64} (9+15 \cos^2\mu + 44 \cos^4\mu-36 \cos^6\mu)
	-\frac{27}{64}\mu (1+2 \cos\mu^2) \tan\mu \nonumber\\
\alpha_2 &= \frac1{128} (3 + 23 \cos^2\mu + 150 \cos^4\mu - 144 \cos^6\mu)
	+\frac1{128} \mu (3 + 24 \cos\mu^2 - 72 \cos^4\mu) \tan\mu \nonumber\\
\beta_0 &=\frac1{64} (111 + 23 \cos\mu^2 + 94 \cos^4\mu - 60 \cos^6\mu) 
	+\frac3{64} \mu (37 + 20 \cos\mu^2 - 4 \cos^4\mu) \tan\mu \nonumber\\
\beta_1 &=-\frac1{64} (27 + 3 \cos\mu^2 + 122 \cos^4\mu - 120 \cos^6\mu)
	-\frac3{64} \mu (9 + 4 \cos\mu^2 - 8 \cos^4\mu) \tan\mu \nonumber\\
\beta_2 &=\frac1{128} (3 - 13 \cos\mu^2 + 234 \cos^4\mu - 240 \cos^6\mu)
	+ \frac3{128} \mu (1 - 4 \cos\mu^2) \tan\mu
\end{align}
Also
\begin{align}
\kappa &= 1 - \frac3{1024} \gamma^2 [74 - 18 (1+2 \cos^2\mu) \cos2\tau
	+ (1 + 8 \cos^2\mu - 24 \cos^4\mu) \cos4\tau ] + O(\gamma^4) \nonumber \\
\lambda &= 1 - \frac3{1024} \gamma ^2 [ 74 +40 \cos^2\mu -8 \cos^4\mu 
	+ 2( -9 -4 \cos^2\mu + 8 \cos^4\mu) \cos2\tau \nonumber \\
       &+ (1-4 \cos^2\mu ) \cos4\tau ] + O(\gamma^4)
\end{align}
and
\begin{align}
\bar a&=\bar\alpha_0(\mu) + \bar\alpha_1(\mu) \cos 2\tau
	+ \bar\alpha_2(\mu) \cos 4\tau \\ \nonumber
\bar b&=\bar\beta_0(\mu) + \bar\beta_1(\mu) \cos 2\tau
	+ \bar\beta_2(\mu) \cos 4\tau
\end{align}
with
\begin{align}
\bar\alpha_0 &= -\frac1{64} \cos^2 \mu (37 -126 \cos^2\mu + 120 \cos^4\mu) 
	\nonumber\\
\bar\alpha_1 &=\frac3{64} \cos^2\mu (3 -44 \cos^2\mu +36 \cos^4\mu) \nonumber\\
\bar\alpha_2 &=-\frac1{128} \cos^2\mu (1 - 222 \cos^2\mu +144 \cos^4\mu) 
	\nonumber\\
\bar\beta_0 &=-\frac1{64} \cos^2\mu (37 -106 \cos^2\mu +60 \cos^4\mu) 
	\nonumber\\
\bar\beta_1 &=\frac1{64} \cos^2\mu (9 -146 \cos^2\mu +120 \cos^4\mu) \nonumber\\
\bar\beta_2 &=-\frac1{128} \cos^2\mu (1 - 234 \cos^2\mu +240 \cos^4\mu)
\end{align}
and
\begin{align}
\mu_0(\tau) = \frac{\pi}{2\kappa(\pi/2,\tau)}
     =\frac{\pi}{2}\left( 1+ \frac3{1024} \gamma ^2 
       (74 - 18 \cos 2\tau + \cos4\tau ) + O(\gamma^4) \right)
\end{align}
The function $\mu_0(\tau)$ has the similar shape to Fig.~\ref{fig10}. We draw the deformation of the 
Penrose diagram in Fig.~\ref{figcom}.
As in section 3.2, the metric can be transformed to the standard BTZ metric
\eqref{btzt} by the coordinate transformation,
\begin{align}
\bar\mu &= \tilde\mu + \frac{3\pi\gamma^2}{512} \sin^2\tilde\mu 
	(9 \cos 2\tilde\tau - 2 \cos4\tilde\tau) + \cdots \nonumber \\
\tau &= \tilde\tau + \frac{3\pi\gamma^2}{1024} (9 \sin2\tilde\mu 
	- \sin4\tilde\mu \cos2\tilde\tau) \sin2\tilde\tau + \cdots
\end{align}
Along the future horizon
\begin{align}
\mu= \tau -\frac{\pi}{2}+ \mu_0\left(\frac{\pi}{2}\right) 
=\tau +\gamma^2 \frac{279 \pi}{2048} + O(\gamma^4)
\end{align}
the horizon length is a monotonically increasing function
\begin{align}
	\mathcal{A}(\tau) = 2\pi R \Big[ 1 +\frac{\gamma^2}{1024} ( &-279 
	+ 93 \cos^2\tau - 782 \cos^4\tau + 3064 \cos^6\tau
      - 4272 \cos^8\tau   \nonumber \\
      & + 1920\cos^{10}\tau 
	 + 279(\frac{\pi}2 - \tau) \tan\tau) +O(\gamma^4)\Big]
\end{align}
Then $\mathcal{A}(\pi/2) = 2\pi R$ which is the BTZ value while $\mathcal{A}(0)$ is given by
\begin{align}
	\mathcal{A}(0) = 2\pi R \left[ 1- \frac{\gamma^2}4 + O(\gamma^4) \right]
\end{align}

A coordinate transformation
\begin{align}
\tau &= \nu - \frac{3\gamma^2}{4096} [
          36 \sin2\nu ( \cos2\sigma + 2\sigma \sin2\sigma )
          + \sin4\nu ( \cos4\sigma + 4\sigma \sin4\sigma ) ] \nonumber \\
\mu &= \sigma + \frac{3\gamma^2}{1024}
	  \sigma (74 + 18 \cos2\nu \cos2\sigma + \cos4\nu \cos4\sigma)
\end{align}
gives the metric of the form \eqref{m0metric} with
\begin{align}
a_\nu &= \frac1{128}[
       222 - 74 \cos^2\sigma + 252 \cos^4\sigma - 240 \cos\sigma^6 \nonumber \\
    &
       + 6 \cos2 \nu (-18 + 3 \cos^2\sigma- 44 \cos^4\sigma + 36 \cos^6\sigma)
         \nonumber \\
       &- \cos4 \nu (-6 + \cos^2\sigma - 174 \cos^4\sigma + 144 \cos^6\sigma)]
       \nonumber \\
a_\sigma &= -\frac1{128}\cos^2\sigma [
       74 - 252 \cos^2\sigma + 240 \cos^4\sigma \nonumber \\
    &
       + 6 \cos2\nu (33 + 44 \cos^2\sigma - 36 \cos^4\sigma) \nonumber \\
     &+ \cos4\nu (-47 + 18 \cos^2\sigma (-7 + 8 \cos^2\sigma) ] \nonumber \\
b_x &= \frac1{512} [
     999 - 8 \cos^2\sigma (7 - 97 \cos^2\sigma + 60 \cos^4\sigma) \nonumber \\
    &
     - \cos2\nu (330 - 16 \cos^2\sigma (15 - 64 \cos^2\sigma
     + 60 \cos^4\sigma)) \nonumber \\
     &- \cos4\nu (-15 + 76 \cos^2\sigma - 960 \cos^4\sigma + 960 \cos^6\sigma)
     ]
\end{align}

\section{Other perturbation with $m^2\neq 0$}\label{appb}
Here we consider only the case with $\ell^2 m^2=3$ ($\Delta=3$) and 
$\kappa_2=0$ in \eqref{legendre} for which 
the explicit form of the solution is given by
\begin{equation} \label{m3h}
	h = \cos^3\mu \left(-\frac23 + \cos^2\tau \right)
\end{equation}
The corresponding solution in the gravity part can be obtained in the
form \eqref{m0sol2} and \eqref{m0sol2ab} 
with
\begin{align}
\alpha_0 &= \frac1{192} (135 -45 \cos^2\mu  - 18 \cos^4\mu +40 \cos^6\mu)
	+ \frac{45}{64} \mu\tan \mu \nonumber\\
\alpha_1 &= -\frac1{576} (165+275 \cos^2\mu -132 \cos^4\mu +108 \cos^6\mu)
	-\frac{55}{192}\mu (1+2 \cos\mu^2) \tan\mu \nonumber\\
\alpha_2 &= -\frac1{1152} (15 +115 \cos^2\mu -402 \cos^4\mu - 144 \cos^6\mu)
	-\frac5{384} \mu (1 + 8 \cos\mu^2 - 24 \cos^4\mu) \tan\mu \nonumber\\
\beta_0 &= \frac1{576} (405 +165 \cos^2\mu  - 94 \cos^4\mu +60 \cos^6\mu)
	+ \frac{5}{192} \mu(27+20 \cos^2\mu  + 4 \cos^4\mu)\tan \mu \nonumber\\
\beta_1 &= -\frac1{576} (165 -115 \cos^2\mu  +118 \cos^4\mu +120 \cos^6\mu)
	- \frac{5}{192} \mu(11-4 \cos^2\mu  + 8 \cos^4\mu)\tan \mu \nonumber\\
\beta_2 &=-\frac1{1152} (15 - 65 \cos\mu^2 + 18 \cos^4\mu - 240 \cos^6\mu)
	- \frac5{384} \mu (1 - 4 \cos\mu^2) \tan\mu
\end{align}
Also
\begin{align}
\kappa &= 1 - \frac5{3072} \gamma^2 
      [54 - 22 (1 + 2\cos^2\mu) \cos2\tau 
      - (1 + 8\cos^2\mu - 24\cos^4\mu ) \cos4\tau ]  + O(\gamma^4) \nonumber \\
\lambda &= 1 - \frac5{3072}\gamma^2 [54 + 40 \cos^2\mu + 8\cos^4\mu 
      - (22 - 8\cos^2\mu + 16\cos^4\mu ) \cos2\tau \nonumber \\
      &-(1 -4 \cos^2\mu ) \cos4\tau ] + O(\gamma^4)
\end{align}
and
\begin{align}
\bar a&=\bar\alpha_0(\mu) + \bar\alpha_1(\mu) \cos 2\tau
	+ \bar\alpha_2(\mu) \cos 4\tau \\ \nonumber
\bar b&=\bar\beta_0(\mu) + \bar\beta_1(\mu) \cos 2\tau
	+ \bar\beta_2(\mu) \cos 4\tau
\end{align}
with
\begin{align}
\bar\alpha_0 &= -\frac1{192} \cos^2\mu (45 +18 \cos^2\mu - 40\cos^4\mu)
	\nonumber\\
\bar\alpha_1 &=\frac1{576} \cos^2\mu (55 + 132 \cos^2\mu -108 \cos^4\mu)
	\nonumber\\
\bar\alpha_2 &=\frac1{1152} \cos^2\mu (5 + 42 \cos^2\mu +144 \cos^4\mu)
	\nonumber\\
\bar\beta_0 &=-\frac1{576} \cos^2\mu (135 +154 \cos^2\mu -60 \cos^4\mu) 
	\nonumber\\
\bar\beta_1 &= \frac1{576} \cos^2\mu (55 +2 \cos^2\mu -120 \cos^4\mu)
	\nonumber\\
\bar\beta_2 &=\frac1{1152} \cos^2\mu (5 -18 \cos^2\mu +240 \cos^4\mu)
\end{align}
and
\begin{align}
\mu_0(\tau) = \frac{\pi}{2\kappa(\pi/2,\tau)}
     = \frac{\pi}{2}\left(1+ \frac5{3072} \gamma ^2 
       (54 - 22 \cos 2\tau - \cos4\tau ) + O(\gamma^4)\right)
\end{align}
We draw the shape of the Penrose diagram on the right side of Fig.~\ref{figcom}. 
The metric can again be transformed to the standard BTZ metric
\eqref{btzt} by the coordinate transformation,
\begin{align}
\bar\mu &= \tilde\mu + \frac{5\pi\gamma^2}{1536} \sin^2\tilde\mu 
	(11 \cos 2\tilde\tau + 2 \cos4\tilde\tau) + \cdots \nonumber \\
\tau &= \tilde\tau + \frac{5\pi\gamma^2}{3072} (11 \sin2\tilde\mu 
	+ \sin4\tilde\mu \cos2\tilde\tau) \sin2\tilde\tau + \cdots
\end{align}
Along the future horizon
\begin{align}
\mu= \tau -\frac{\pi}{2}+ \mu_0\left(\frac{\pi}{2}\right) 
=\tau +\gamma^2 \frac{125 \pi}{2048} + O(\gamma^4)
\end{align}
the horizon length becomes again a monotonically increasing function
\begin{align}
	\mathcal{A}(\tau) = 2\pi R \Big[ 1 +\frac{\gamma^2}{3072} ( &-375
	+ 125 \cos^2\tau +50 \cos^4\tau -264 \cos^6\tau
      +848 \cos^8\tau   \nonumber \\
      & -640\cos^{10}\tau 
	 + 375(\frac{\pi}2 - \tau) \tan\tau) +O(\gamma^4)\Big]
\end{align}
Then $\mathcal{A}(\pi/2) = 2\pi R$ which is the BTZ value as before
while $\mathcal{A}(0)$ is given by
\begin{align}
	\mathcal{A}(0) = 2\pi R \left[ 1- \frac{\gamma^2}{12} + O(\gamma^4) \right]
\end{align}

A coordinate transformation
\begin{align}
\tau &= \nu - \frac{5\gamma^2}{12288} [
          44 \sin2\nu ( \cos2\sigma + 2\sigma \sin2\sigma )
          - \sin4\nu ( \cos4\sigma + 4\sigma \sin4\sigma ) ] \nonumber \\
\mu &= \sigma + \frac{5\gamma^2}{3072}
	  \sigma (54 + 22 \cos2\nu \cos2\sigma - \cos4\nu \cos4\sigma)
\end{align}
gives the metric of the form \eqref{m0metric} with
\begin{align}
a_\nu &= \frac1{1152}[
	6(135 - 45 \cos^2\sigma - 18 \cos^4\sigma + 40 \cos\sigma^6) \nonumber \\
    &
       + \cos2\nu (-660 + 110 \cos^2\sigma +264 \cos^4\sigma -216 \cos^6\sigma)
       \nonumber \\
       &+ \cos4\nu (-30 + 5\cos^2\sigma + 282 \cos^4\sigma + 144 \cos^6\sigma)]
       \nonumber \\
a_\sigma &= -\frac1{1152}\cos^2\sigma [
       6(45 + 18 \cos^2\sigma - 40 \cos^4\sigma ) \nonumber \\
    &
     + 2 \cos2\nu (605 - 132 \cos^2\sigma + 108 \cos^4\sigma) \nonumber \\
     &+ \cos4\nu (235 - 18 \cos^2\sigma (29 + 8 \cos^2\sigma) ] \nonumber \\
b_x &= \frac1{4608} [
     3885 + 120 \cos^2\sigma - 872 \cos^4\sigma + 480 \cos^6\sigma \nonumber \\
    &
     - 2 \cos2\nu (975 +8 \cos^2\sigma (-125 + 44 \cos^2\sigma
     + 60 \cos^4\sigma)) \nonumber \\
    &- \cos4\nu (75 - 380 \cos^2\sigma + 192 \cos^4\sigma - 960 \cos^6\sigma)
     ]
\end{align}

\end{document}